\documentclass[conference]{IEEEtran}
\makeatletter
\IEEEoverridecommandlockouts
\usepackage{balance}
\usepackage{algorithm} 
\usepackage{eucal}
\usepackage{amssymb}
\usepackage[algo2e]{algorithm2e}

\usepackage{cite}
\usepackage{amsmath,amssymb,amsfonts}
\usepackage{graphicx}
\usepackage{xcolor,soul}
\usepackage[flushleft]{threeparttable}
\usepackage[caption=false]{subfig}
\usepackage{booktabs}
\def\BibTeX{{\rm B\kern-.05em{\sc i\kern-.025em b}\kern-.08em
	T\kern-.1667em\lower.7ex\hbox{E}\kern-.125emX}}
\usepackage{algpseudocode}
\usepackage{bm}
\usepackage{array}
\usepackage{multirow}
\usepackage[draft,footnote,nomargin]{fixme}

\begin{document}
\IEEEoverridecommandlockouts
\title{
Segmented Learning for Class-of-Service\\ Network Traffic Classification

	\thanks{This work is supported in part by the Ontario Center of Innovation (OCI) ENCQOR 5G development program and the Natural Sciences and Engineering Research Council of Canada (NSERC) under Grant RGPIN-2020-04661.}
	\thanks{(\textit{Corresponding author: Xiao-Ping Zhang}, e-mail: xzhang@ryerson.ca)}
}

\author{\IEEEauthorblockN{ Yoga Suhas Kuruba Manjunath$^\ddag$, Sihao Zhao$^\ddag$, Hatem Abou-zeid$^*$,\\ Akram Bin Sediq$^*$, Ramy Atawia$^*$, Xiao-Ping Zhang$^\ddag$
	}
	\IEEEauthorblockA{$\ddag$Department of Electrical, Computer and Biomedical Engineering, Ryerson University, Toronto, ON, Canada \\
		*Ericsson Canada, Ottawa, Canada}\\
}

\maketitle

\begin{abstract}

Class-of-service (CoS) network traffic classification (NTC) classifies a group of similar traffic applications. The CoS classification is advantageous in resource scheduling for Internet service providers and avoids the necessity of remodelling. Our goal is to find a robust, lightweight, and fast-converging CoS classifier that uses fewer data in modelling and does not require specialized tools in feature extraction. The commonality of statistical features among the network flow segments motivates us to propose novel segmented learning that includes essential vector representation and a simple-segment method of classification. We represent the segmented traffic in the vector form using the EVR. Then, the segmented traffic is modelled for classification using random forest. Our solution's success relies on finding the optimal segment size and a minimum number of segments required in modelling. The solution is validated on multiple datasets for various CoS services, including virtual reality (VR). Significant findings of the research work are i) Synchronous services that require acknowledgment and request to continue communication are classified with 99\% accuracy, ii) Initial 1,000 packets in any session are good enough to model a CoS traffic for promising results, and we therefore can quickly deploy a CoS classifier, and iii) Test results remain consistent even when trained on one dataset and tested on a different dataset. In summary, our solution is the first to propose segmentation learning NTC that uses fewer features to classify most CoS traffic with an accuracy of 99\%. The implementation of our solution is available on GitHub.

\end{abstract}
%

\section{Introduction}
\label{sec:intro}

The surge in Internet traffic has posed a significant challenge for Internet service providers (ISPs) to strategize the resource allocation for a better quality of service (QoS) \cite{itc}. The ISPs use network traffic classification (NTC) to understand the traffic \cite{finsterbusch2013survey} and decide on network management. The conventional multi-class NTC \cite{8026581, 8647128, 8651277, ysk} classifies the network traffic into respective applications such as Facebook, Netflix, YouTube traffic, etc. Conventional NTC needs to be remodelled for every new application type released on the Internet, which is expensive in training and adds up to the maintenance cost for ISPs \cite{pacheco2018towards,ZHAO202122}. Unlike the conventional multi-class NTC, the class-of-service (CoS) classifier groups similar applications and maps them into one CoS type. For example, Skype, Hangouts, and Facebook chat services are treated as ``chat" CoS types instead of three different classes. The CoS classification is more suitable and offers less latency in scheduling for better QoS \cite{10.1109/MCOM.2009.4785383}. The CoS NTC is getting more attention from researchers to provide better solutions because of its necessity in faster resource scheduling for a better QoS \cite{mehta2017survey}.

Interactive (e.g., VoIP), bulk data transfer (e.g., FTP), streaming (e.g., video), and transactional (e.g., chat) CoS types are investigated in \cite{roughan2004class}. The linear discriminant analysis and the K-nearest neighbours-based solutions achieve 43\% and 86\% accuracy, respectively. The solution has a complex feature engineering process, which uses packet-level, flow-level, connection-level, intra-flow level, and multi-flow level features to obtain statistics for CoS classification, and the performance is not satisfactory. Shapley value-based super-features are used to represent the CoS traffic data in \cite{chowdhury2019explaining}. The preprocessing requires 266 features to obtain the Shapley values. The work uses a neural network in classification. Currently, the solution stands as the state-of-the-art (SOA) solution with an accuracy of 92.5\%. However, the result is validated on only 10\% of the dataset. No works consider the network traffic in segments for classification to take advantage of the commonality of statistical features in traffic features.

Based on the literature review, we observe the following issues in the current CoS classifiers: i) complex feature engineering methods that are impractical in a real-world scenario, ii) less robust solutions evaluated on a limited dataset, iii) an extensive number of packets in modelling, iv) complex classifiers with huge training complexity and v) lack of taking advantage of properties of traffic segments. The research gaps motivate us to find a simple feature engineering technique that is feasible in a real-world scenario, and a quicker method to model a robust CoS classifier that learns the CoS services holistically to avoid retraining.

We propose a novel segmented learning that includes an essential vector representation (EVR) and a simple-segment method of classification (S2MC) algorithms. Segmented learning treats the network traffic in segments and learns the unique and common properties among the segments pertaining to a CoS class. The traffic packets are collected with four primary features: packet direction, inter-arrival time, packet length, and timestamp segments. The EVR forms one feature vector per segment with eleven elements: packet size statistics, inter-arrival time statistics, the number of packets in the uplink and downlink directions, and time duration from the segment. The main aim of segmented learning is to find the ideal segment size ($N$) so that the statistical feature shows less or no variation, and a minimum number of segments ($S_T$) is required for modelling a classifier. Therefore, segmented learning involves finding ($N, S_T$) pair by employing a self-learning heuristic technique that uses a random forest. The solution is validated on different datasets \cite{data,zhao2021vr,draper2016characterization} that consist of several CoS types such as file transfer, remote cloud service, video, VoIP, virtual reality (VR), chat, audio, and peer to peer (P2P). Hence, we provide a robust CoS classifier. In addition, we show the generalization of the solution by training on one dataset and testing on a different dataset.

The implementation of our work is available in GitHub \cite{code}. A few crucial findings of the work are as follows. i) Synchronous services (e.g. video, VR streaming and many others) that requires acknowledgements from the other end are modelled with 99\% accuracy. ii) Test accuracy is consistent even when trained on one dataset and testing on the other dataset for synchronous services. iii) We find that $N$=20 and $S_T$=50 provide best performance for both datasets used in the work. iv) The maximum packet length, maximum mean, and standard deviation of inter-arrival time consistently provide more information for classification using random forest. v) Asynchronous services (e.g. Chat, email and many others) are not modelled accurately when delay occurs in communication. vi) Different parameters such as user inactivity, delay in networks or congestion are not effectively captured in segmented learning. We will discuss more on synchronous and asynchronous services in later sections.


The contributions of the paper are as follows.
\begin{itemize}
    \item We develop a lightweight CoS classifier that uses fewer features captured by simple sniffer tools like Wireshark or TCPdump.
    \item  The new segmented learning, which includes EVR and S2MC, has a superior classification compared to state-of-the-art solution \cite{chowdhury2019explaining}. 
    \item  We prove the robustness of the solution by validating on various CoS traffic services (virtual reality, video, chat, and many others) and on multiple datasets.
    \item To the best of the author's knowledge, ours is the first lightweight, robust, and high-performing CoS classifier.
\end{itemize}

\label{subsec:evr}
\begin{figure}
	\centering
	\includegraphics[trim=0 0 0 0 ,width=\linewidth]{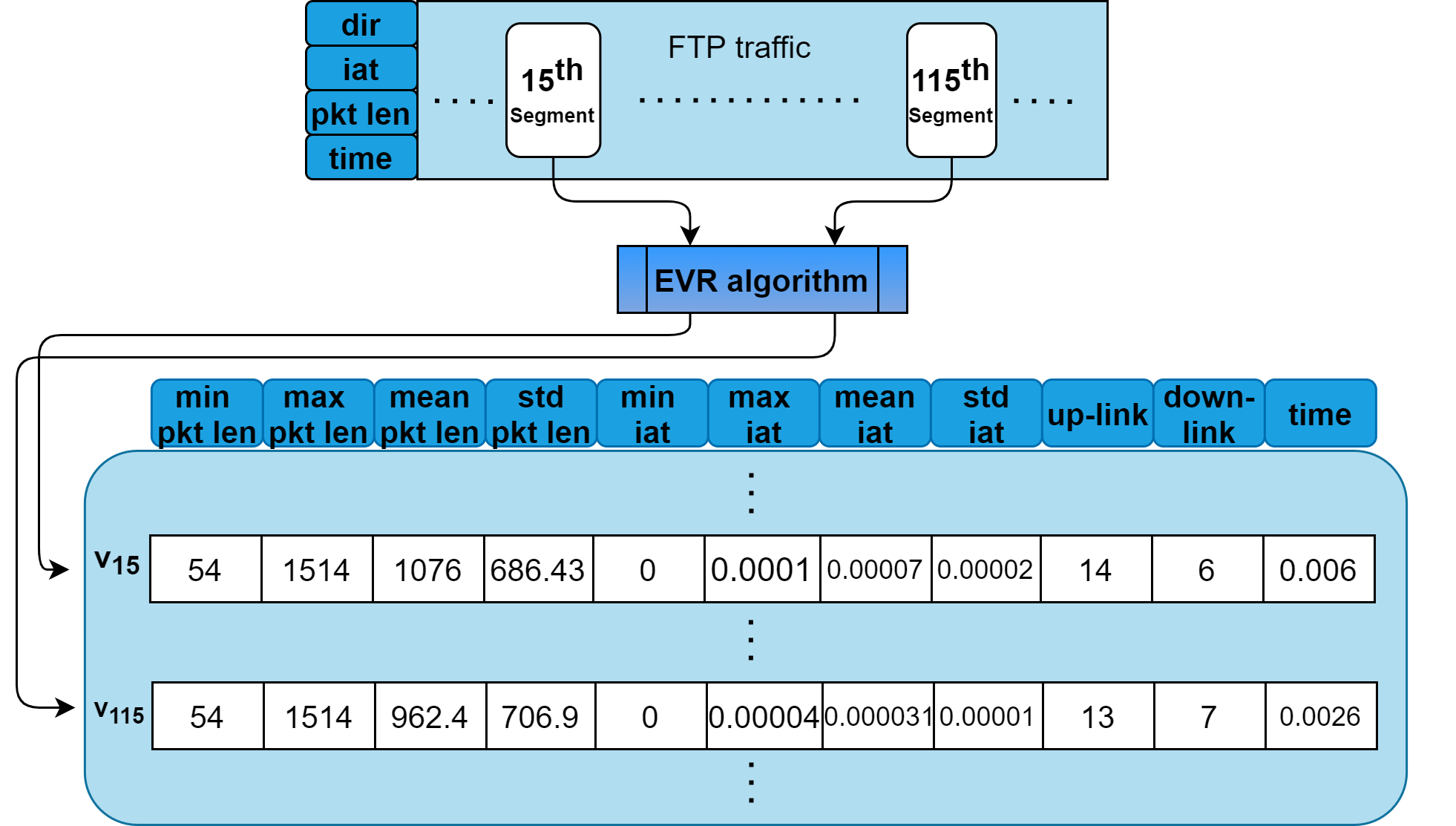}
	\caption{Representing $\bm{v}_n$ for the $n$\textsuperscript{th} segment. We show the representation of EVR for 15\textsuperscript{th} and 115\textsuperscript{th} segment from FTP.}
	\label{fig:evr}
\end{figure}

\section{Segmented learning}

Segmented learning treats the traffic in segments. Then, statistics of segments are modelled for classification. The fruition of segmented learning depends on finding an optimal segment size $N$ and a minimum number of segments $S_T$ required for modelling. We introduce the EVR and S2MC to find segment size and number of segments pair ($N, S_T$) for successful classification of CoS traffic.

\subsection{Essential Vector Representation (EVR)}

\begin{algorithm}
\caption{EVR Algorithm }\label{alg:alg1}
\KwData{total number of segments required ($S_T$), segment size ($N$) and Raw traffic data with time, pkt len, iat, dir.}
\KwResult{$\bm{S}$}
 initialization\;
 
 $\bm{S}$ = $[$ $]\;$ 
 
 $n = 0$
  
\While{$n$ != $S_T$}{
 
    initialize data\_packets and time\;
    
    \While{length(data\_packets) != segment size ($N$)}{
    data = collect data\_packets\; 
    
    time = timer value}\;
    
    $\bm{v}_n$ = stats(data.pktlen, data.iat, data.up,  data.dn, time) \;
    
    $\bm{S}.append(\bm{v}_n)\;$
    
    $n++\;$

  }
\Return {$\bm{S}$}
\end{algorithm}

In the EVR, traffic segments are represented in a vector form, denoted by $\bm{v}_n$, where $n$ represents the index of the vector. The process of the vector formation using EVR is illustrated in Fig. \ref{fig:evr}. Every segment consists of a $N$ number of packets with 4 features in each, i.e., packet direction, inter-arrival time, packet length, and timestamp. The vector $\bm{v}_n$ consists of 11 features derived from the above 4 features. The first 4 are the packet length's statistical information (minimum, maximum, average, and standard deviation derived from pkt len in the figure). The second 4 features are the statistical information of the packet inter-arrival time (iat in the figure). The next feature is the number of packets in uplink and downlink directions. Finally, the last feature is the time taken to form the traffic segment. In total, eleven features are used in the proposed method, as shown in Fig. \ref{fig:evr}.

Algorithm \ref{alg:alg1} shows the EVR implementation that is invoked by the S2MC algorithm, which will be explained in the next section. Algorithm \ref{alg:alg1} requires the segment size ($N$), the number of segments required ($S_T$) and the raw traffic data with the basic features. EVR algorithm perform segmentation and vectorization based on the input. Every vector $\bm{v}_n$ is stacked to form a matrix $\bm{S}$. Stacking action is performed by “append” in Algorithm \ref{alg:alg1}.

\begin{figure}
	\centering
	\includegraphics[width=\linewidth]{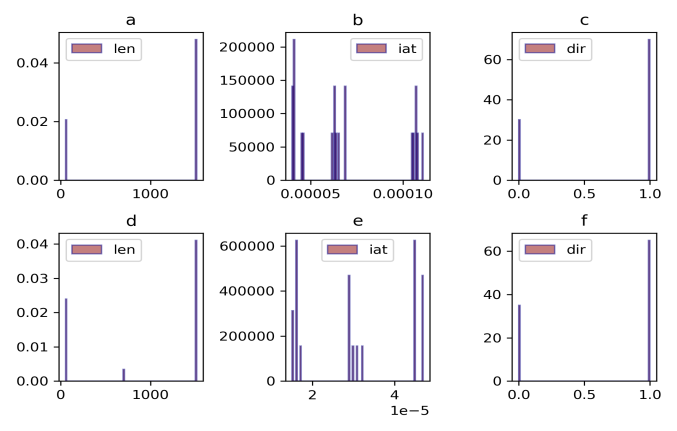}
	\caption{Histogram of packet length (len), inter-arrival time (iat), and direction (dir) for file transfer service. (a,b,c) shows histogram for the 15\textsuperscript{th} segment with $N$=20 packets. (d,e,f) shows histogram for the 115\textsuperscript{th} segment with $N$=20 packets. Histograms at different parts of traffic look similar, showing that the statistical information of segments can be used to represent the traffic data.}
	\label{fig:ftp}
\end{figure}

We analyze the rationality behind the 11 statistical features present in $\bm{v}_n$. Asynchronous services such as chat and VoIP use smaller packets than synchronous services such as video and file transfer. Because the service type and quality of service are essential in synchronous services. Therefore, statistics of packet size provide unique information on different CoS. Inter-arrival time is another critical feature. The delay-sensitive services like VoIP and chat show aggressive packet bursts in both directions and smaller inter-arrivals. However, if the users are inactive or take a long time to respond, the inter-arrival time might be more significant in those services. Video or file transfer services use the cache and spend mechanism \cite{7987775}, which causes a slightly longer inter-arrival time because the server transmits data in bulk. The client caches the data in a buffer. Once the buffer gets depleted to lower than the threshold, it sends a request for the remaining data to the server. However, the inter-arrival time will be less in those data transferred in bulk, even for file transfer or video streaming services. We capture this behavioural information well in the statistics from the traffic segments, and we observe that these characteristics remain almost constant if we choose the segment size $N$ correctly. We can observe in Fig. \ref{fig:ftp} that statistical information of packet length, inter-arrival time and direction is almost the same in the 15\textsuperscript{th} and 115\textsuperscript{th} segments for $N$=20 for file transfer service. However, we see the differences in statistical information of features for different values of $N$. How to select the value of $N$ will be explained in the next subsection. Therefore, the 11-feature vector with statistical information can provide a good representation of different CoS types.

\subsection{Simple-Segment Method of Classification (S2MC)}

The fruition of the solution majorly depends on finding the optimal segment size (number of packets in a segment) $N$. That means we need to find segment sizes that produce less or no variation in statistical information. The statistical information can vary for a given CoS traffic at different stages in a session. The minimum number of segments $S_T$ required for modelling the CoS traffic is also an imperative parameter. With an appropriate $S_T$, we can save time in modelling. Because of the high randomness of CoS traffic, we employ the S2MC that employs a heuristic machine learning method to find the $S_T$ value. We observe variation in the statistical information at the beginning of the session due to irregular packet transmissions related to protocol negotiations. However, that is unique to the CoS types, and statistical information remains constant throughout the session. Intuitively, the solution can perform well if we choose the packets at the beginning and a few packets from the middle part of a session. We can use irregularities and stationarity at the beginning and middle of the session, respectively.

The $N$ and $S_T$ are positive integers that can take values from 0 to $\infty$. At this stage, it is difficult to find the values analytically because of randomness and hence, we employ a heuristic method to take advantage of machine learning. The S2MC chooses different values from the pool $N=\{10,20,\cdots,50\}$ and $S_T=\{10,20,\cdots,50\}$. Any value less than 10 is too low for $N$ because segment information might be statistically insignificant, and we choose 50 as the maximum to not exhaust the data. A small value for $S_T$ might underfit, and a huge value might overfit. Therefore, the S2MC uses the random forest classifier to evaluate the performance and optimize the algorithm for a given network with a given type of CoS. The ($N$, $S_T$) pair might not be universal and depends on the network's behaviour and health. However, the found pair value by Algorithm \ref{alg:alg2} holds good for all CoS types within a network. In our experiment, the ($N$, $S_T$) remains the same for both datasets.

\begin{algorithm}
\caption{S2MC model and segment size selection}\label{alg:alg2}
\KwData{CoS traffic data, benchmarkAccuracy}
\KwResult{final model, $N$}
 initialization\;
 
  
 \For{$S_T=10$; $S_T\leqslant50$; $S_T\mathrel{+}=10$}{
    \For{$N_f=10$; $N_f\leqslant50$; $N_f\mathrel{+}=10$}{
     trainData = EVR$\left(S_T,N_f,CoS \;data\right)$
     
     testData = EVR$\left((L-S_T),N_f,CoS \;data \right)$
     
     model = RandomForest(10,trainData)
     
     performanceMetrics = model.test(testData)
     
     \If{performanceMetrics $>$ benchmarkAccuracy}{
        $N$ = $N_f$
        
        finalModel = model
        
        break
     }
    }
  }
\Return {final model, $N$}
\end{algorithm}

\begin{figure}
	\centering
	\includegraphics[width=\linewidth]{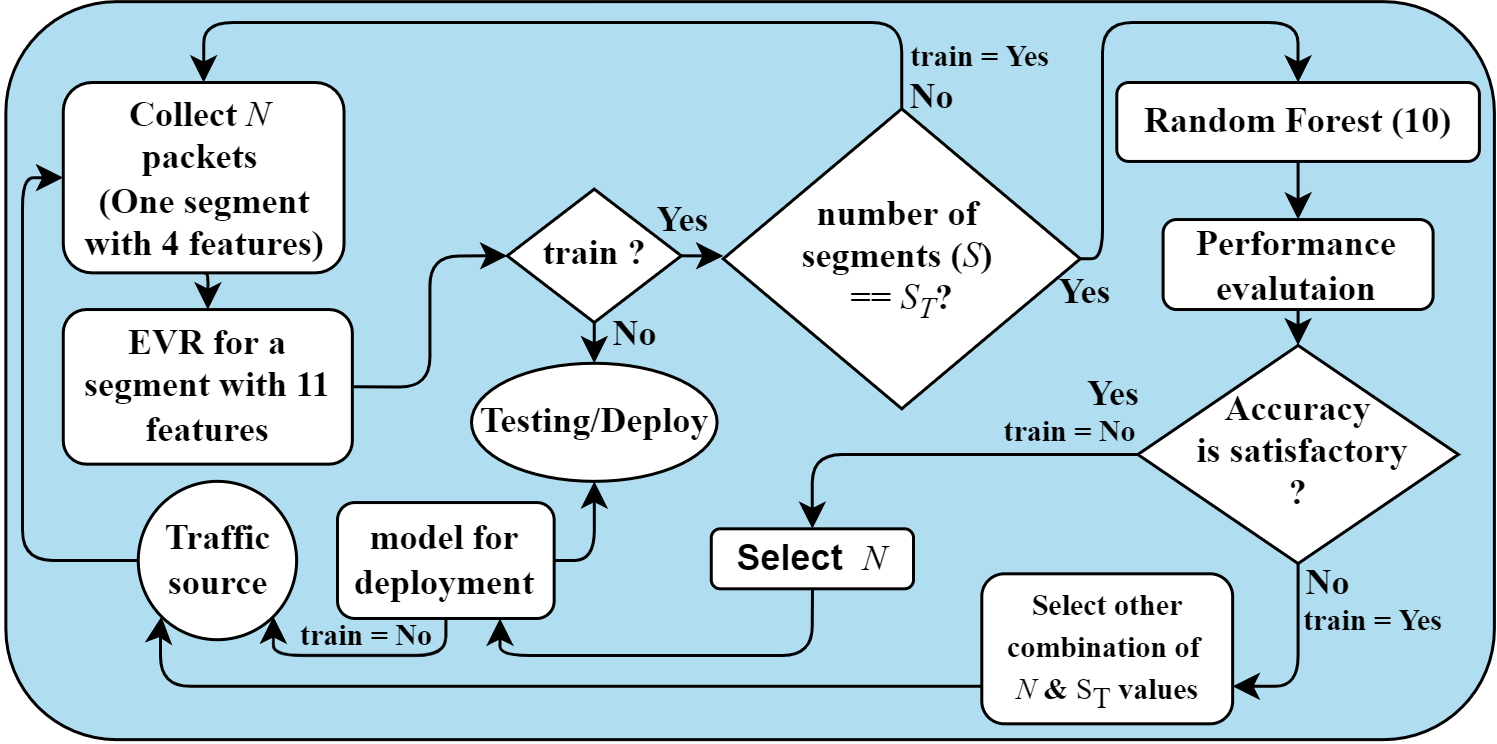}
	\caption{Flow chart of the segmented learning that includes the EVR and S2MC}
	\label{fig:s2mc_t}
\end{figure}

Algorithm \ref{alg:alg2} shows the model and segment size selection using S2MC. $L-S_T$ represent the size of the testing data. $L$ is the size of the original data. The algorithm employs optimization for every number from the $N$ and $S_T$ pool to find the best pair. Random forest is trained based on the selected $S_T$ number of segments, and performance is analyzed on the new data. We use 97\% as the benchmark accuracy for Dataset I and 99\% for Dataset II. The benchmark accuracies are selected arbitrarily. If it is lower than the benchmark accuracy, the algorithm selects new values from the $N$ and $S_T$ pool and continues until the algorithm converges. Once the algorithm converges, we keep the $N$ value required for the deployment and trained model on the $S_T$ number of segments. The general flow of segmented learning is shown in Fig. \ref{fig:s2mc_t} that consists of EVR and S2MC.

\section{Experimental Setup \& Results}



The experiments are done on a computer with Intel i9 processor, 32 GB RAM and Nvidia RTX 2080S GPU with 8 GB RAM. The required software is implemented in python using TensorFlow and Keras APIs along with sci-kit learn libraries. 
\subsection{Dataset Preparation} \label{datapre}


We use the datasets from \cite{data}, \cite{zhao2021vr}  and \cite{draper2016characterization}. We combine the datasets from \cite{data} and \cite{zhao2021vr} to form Dataset I, which consists of 16 different services and 850,857 traffic packets in total. Dataset II from \cite{draper2016characterization} provides the traffic captures of 19 different services with 6,977,991 packets in total. Both datasets represent the real-world traffic. We use four features, namely timestamps, packet size, uplink and downlink inter-arrival time, and packet direction as shown in Fig. \ref{fig:evr}. It does not require scanning the contents of the packets to obtain those four features. Therefore, our method does not breach privacy, and simple sniffers like Wireshark \cite{lamping2004wireshark} or TCPDump \cite{fuentes2005ethereal} are good enough. Moreover, we can obtain the packet captures at any point in the network. Our solution is the first to use only 4 features in traffic classification problems to the best of our knowledge. We combine the similar applications and label them with CoS types. The services and labels of Dataset I and Dataset II are shown in Table \ref{table:CoS labelling of Dataset I}.

\begin{table}
\caption{CoS Labelling of Dataset I and Dataset II}
\centering 

\begin{tabular}{
>{\centering\arraybackslash}m{0.2cm} 
>{\centering\arraybackslash}m{1.3cm} 
>{\centering\arraybackslash}m{1.5cm}
>{\centering\arraybackslash}m{4cm}} 
\hline
 &CoS label &service type & services \\
\hline 
\multirow{5}{4em}{\rotatebox[origin=c]{90}{Dataset I}} & File transfer& Synchronous & FTP, OneDrive\\
&Video& Synchronous & RTP, YouTube, UDP video\\
&VoIP& Synchronous & JoinMe, line, Skype, Zoom\\
&Remote cloud& Synchronous & ncuCloud, RDP, VmWare, Xen\\
&VR& Synchronous & cloud and local gaming\\
\hline
\hline
\multirow{7}{4em}{\rotatebox[origin=c]{90}{Dataset II}} & File transfer& Synchronous & FTP, SCP etc.\\
&Video & Synchronous& Vimeo, Youtube, Netflix \\
&VoIP& Synchronous & Facebook voice, Skype voice, etc.\\
&Chat& Asynchronous & Facebook chat, ICQ etc.\\
&Audio& Asynchronous & Spotify\\
&P2P& Asynchronous & Bittorrent\\
&Email& Asynchronous & SMPTS \\
\hline
\end{tabular}
\label{table:CoS labelling of Dataset I}
\end{table}

\subsection{Performance Metrics}


We use accuracy, recall, precision, F1 score and false-negative rate (FNR) as the performance metrics. The equation of all performance metrics are given below:

\begin{equation}
    Accuracy = \frac{TP+TN}{TP+TN+FP+FN},\nonumber
\end{equation}
\begin{equation}
   Recall = \frac{TP}{TP+FN},\nonumber
\end{equation}
\begin{equation}
   Precision = \frac{TP}{TP+FP},\nonumber
\end{equation}
\begin{equation}
    F1 =    2\times \frac{Precision\times Recall}{Precision+Recall}, \nonumber
\end{equation}
\begin{equation}
    FNR=\frac{FN}{TP+FN}\nonumber
\end{equation}
where, FP is the false-positive that signifies the flows when the flow is not present in reality. False-negative (FN) represents no detection when the flows present in reality. True-negative (TN) shows correct no detection. The correct detection is captured in true-positive (TP). The false-negative rate (FNR) in the NTC problem is important because we need to know when our model indicates no traffic, whereas the traffic is present in reality. Generally, a model with a higher F1 score is robust and reliable. A model with lower FNR is better in classification. Ideally, a model with a higher F1 score and lower FNR is preferred.

\subsection{Performance on the Real-world Datasets}
\label{pm}

\begin{figure}[htp]
\centering
\subfloat[]{%
  \includegraphics[clip,width=0.8\columnwidth]{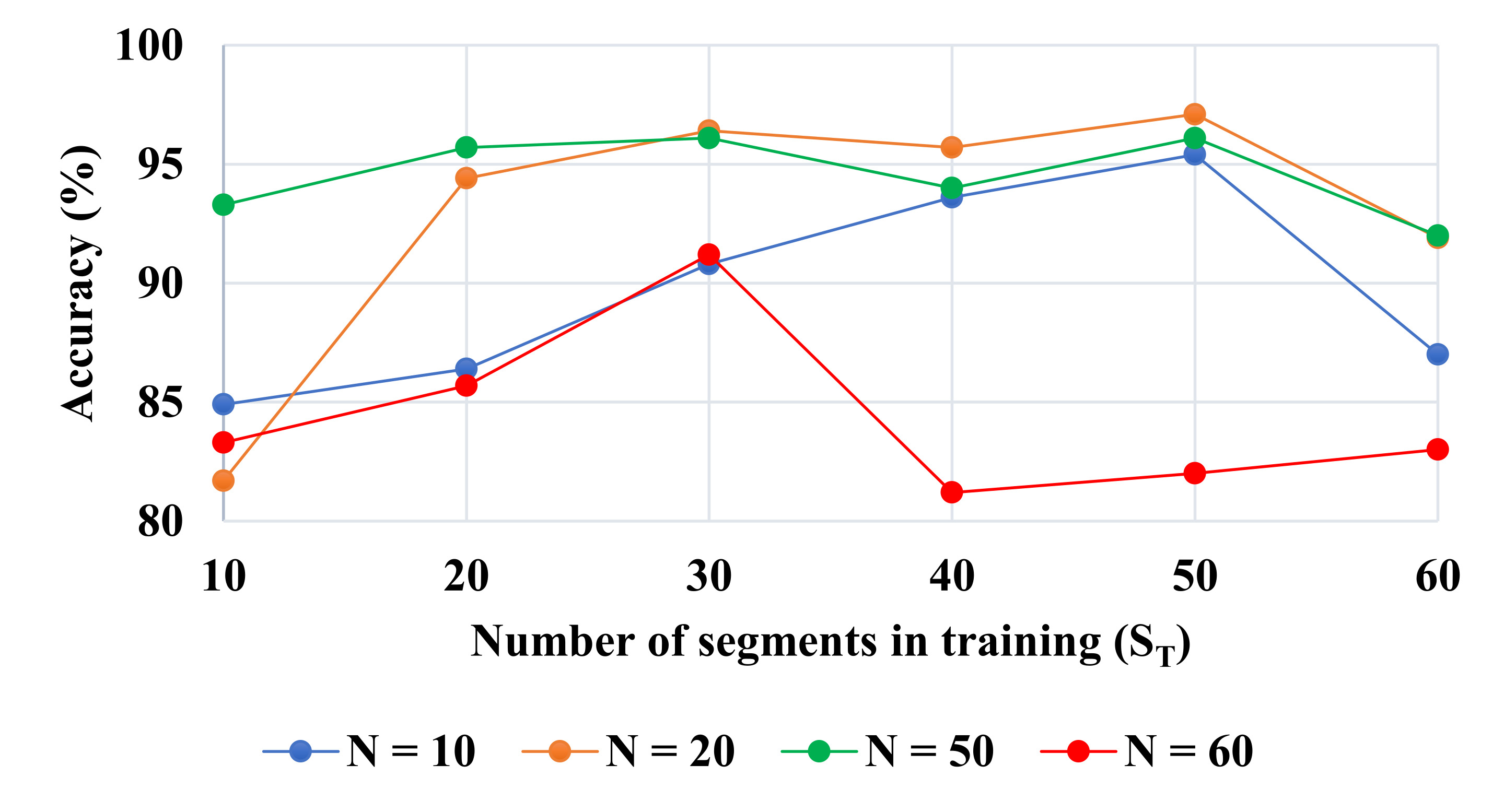}%
}
\vspace{0.1cm}
\subfloat[]{%
  \includegraphics[clip,width=0.8\columnwidth]{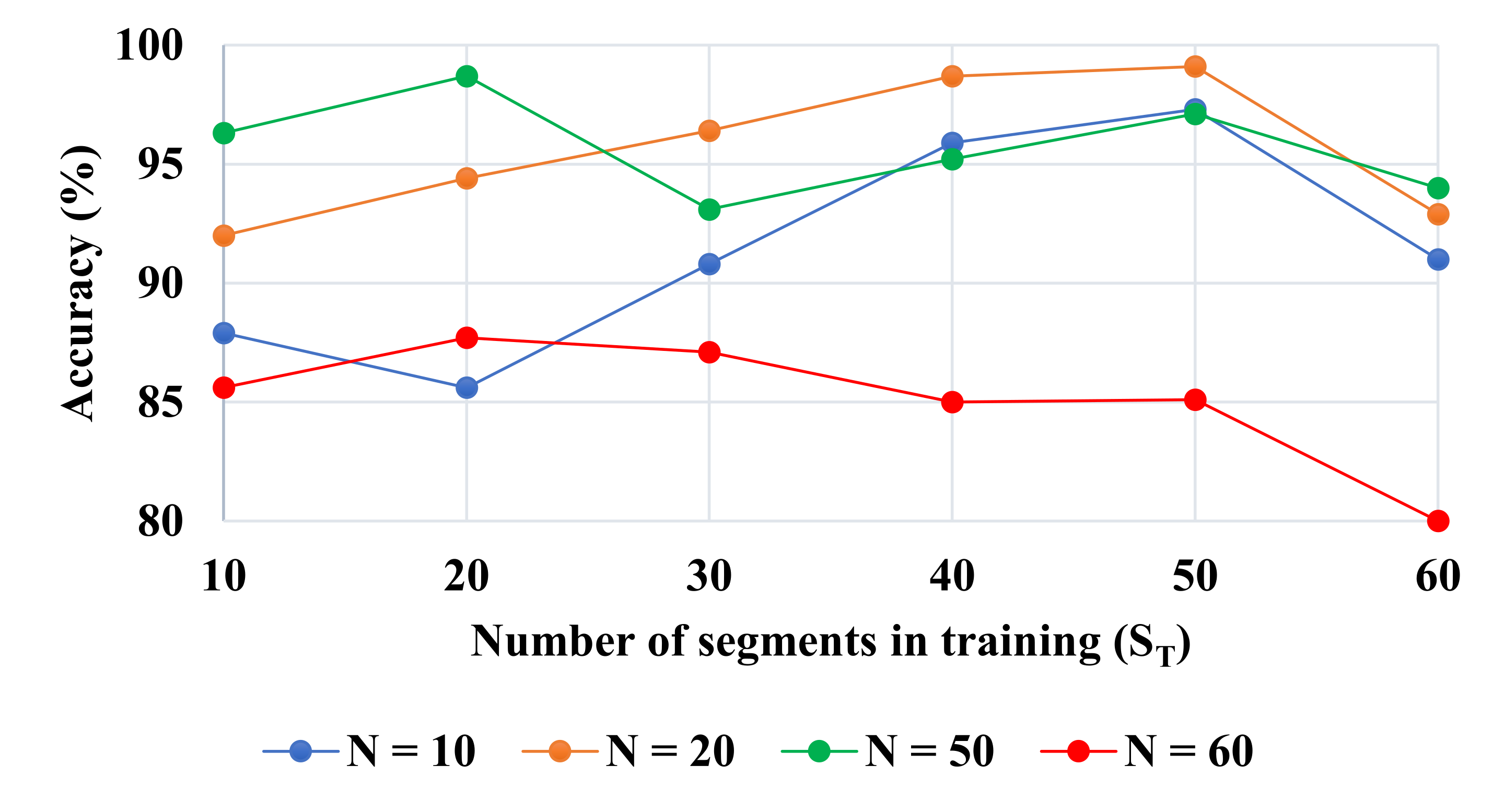}%
}
\caption{Effects of different values of $N$ and $S_T$ on Dataset I (a) and Dataset II (b) modelling. We find the best accuracy when $N$=20 and $S_T$=50 in both cases.}
\label{fig:NS}
\end{figure}

We perform three different experiments. Dataset I is used for both training and testing in the first experiment. In the second experiment, for training and testing Dataset II is used. Finally, in the third experiment, Dataset II is used in training and Dataset I in testing. The performance of the solution for different values of $N$ and $S_T$ is shown in Fig. \ref{fig:NS} for both datasets. We observe 96.1\% accuracy when $N$=50 and $S_T$=30 for Dataset I from Fig \ref{fig:NS} (a). Values greater than $S_T$=30 provide lower accuracy for $N$=50. The results show that with greater $N$ values, segmented learning cannot capture the holistic characteristics required for classification. The observation is similar on Dataset II. Therefore, we provide results for $N$=20 and $S_T$=50 and these values are consistent in all the three experiments. $S_T$ greater than 50 shows poor performance mainly because of data exhaustion.

\begin{figure}
	\centering
	\includegraphics[width=\linewidth]{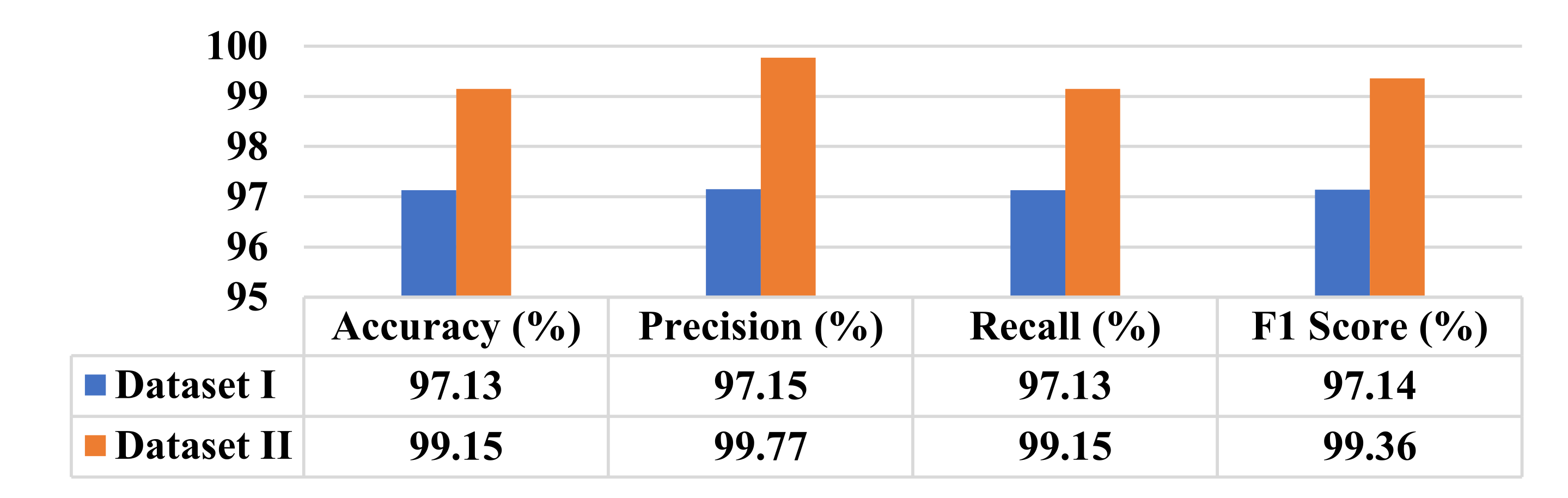}
	\caption{Results of performance metrics on test set of Dataset I and Dataset II}
	\label{fig:d2_p}
\end{figure}

Our solution shows an accuracy of 97.13\% and an F1 score of 97.14\% in experiment 1 as given by Fig. \ref{fig:d2_p}. In the second experiment, we obtain 99.15\% accuracy and 99.36\% F1 score as given by Fig. \ref{fig:d2_p}. We notice that Dataset II is highly unbalanced, which is one of the reasons for miss-classifications. The number of test segments, accuracy per class, and FNR in both experiments are given in Table \ref{table:Accuracy and FNR for dataset I}. Classes like FTP, Video, and VoIP are modelled perfectly in both experiments as given by the table. Email and P2P services show FNR of 46\% and 42\%, respectively. We observe the variations in statistical information among the segments from these services because of asynchronous timeout in the protocols. 

In the third experiment, we use FTP, video, and VoIP CoS type from Dataset II in training and test the same CoS types but from Dataset I. We notice 99.8\% accuracy and 99.11\% F1 score in that case. As expected, synchronous communication CoSs such as video, VoIP and VR, have better results which is observed in the third experiment.







\begin{figure*}
	\centering
	\includegraphics[width=0.9\linewidth]{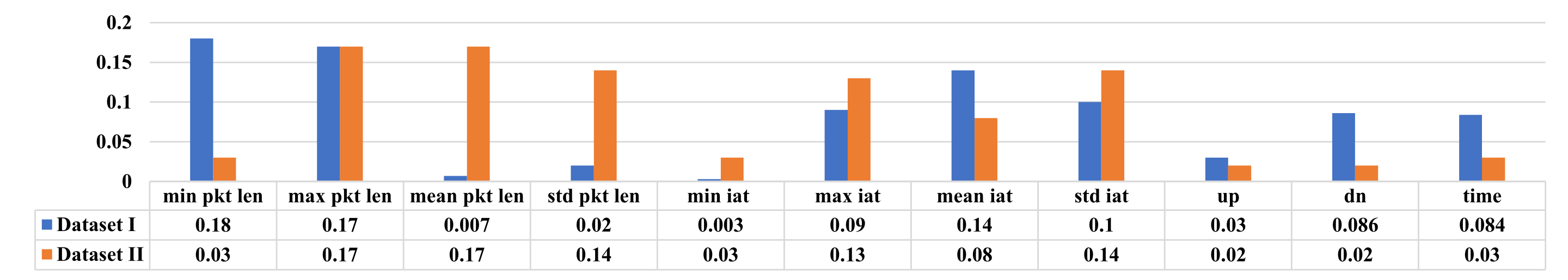}
	\caption{Feature importance in the S2MC classifier}
	\label{fig:fi}
\end{figure*}

\subsection{Feature Importance}


The essential features that provide information in the random forest classifier are shown in Fig. \ref{fig:fi} for both datasets. The impurity-based feature importance is calculated using the Scikit-learn library \cite{scornet2020trees}. In the figure, 0 represents the minimum contribution, and 1 represents the maximum contribution from the features towards classification. We can see from Fig. \ref{fig:fi} that, in Dataset I, min. and max. packet length provides more information than the mean and std of the packet length. Also, max., mean, and std of inter-arrival time, the number of downlink packets, and time duration all together provide good information in classification. In Dataset II, max., mean, and std of packet length provide more information. In addition, max., mean, and std of inter-arrival time also provide good information in classification.


\begin{table}
\caption{Accuracy and FNR for Dataset I and Dataset II}
\centering

\begin{tabular}{
>{\centering\arraybackslash}m{0.2cm} 
>{\centering\arraybackslash}m{2cm} 
>{\centering\arraybackslash}m{2cm}
>{\centering\arraybackslash}m{1cm} 
>{\centering\arraybackslash}m{1cm}}
\hline
&CoS label & Number of test segments & Accuracy (\%) & FNR \\
\hline 
\multirow{5}{4em}{\rotatebox[origin=c]{90}{Dataset I}} &File transfer & 9849 & 98.07 & 0.04\\
&Video & 5153 & 99.89 & 0.0005\\
&VoIP & 12948 & 99.54 & 0.014\\
&Remote cloud & 4202 & 98.47 & 0.07\\

&VR & 10143 & 98.28 & 0.02\\
\hline
\hline
\multirow{7}{4em}{\rotatebox[origin=c]{90}{Dataset II}} 
&File transfer&	286971&	99.86&	0.001\\
&Video&	6504&	99.96&	0.02\\
&VoIP&	49817&	99.85&	0.009\\
&Chat&	647&	99.82&	0.19 \\

&Audio&	54&	99.44&	0\\

&P2P&	3803&	99.52&	0.42\\

&Email&	757&	99.78&	0.46\\
\hline
\end{tabular}
\label{table:Accuracy and FNR for dataset I}
\end{table}





\begin{table}
  \caption{Accuracy and FNR comparison of our solution with prior work.}
  \label{table:FNR comparision}
  \centering
  \begin{tabular}{ccccccc}
    \toprule
    \multirow{2}{*}{CoS class} &
      {SOA work\cite{chowdhury2019explaining}} &
      {Our solution}\\
      & {FNR} & {FNR} \\
      \midrule
    File transfer & 0.27 & 0.001\\
    Video & 0.22 & 0.02\\
    VoIP & 0.36 & 0.009 \\
    chat & 0.35 & 0.19\\  
    Audio & 0.27 & 0\\
    P2P & 0.04 & 0.42\\
    Email & 0.14 & 0.46\\
    \hline
Accuracy (\%) & 92.5 & 99.13\\
    \bottomrule
  \end{tabular}
\end{table}

\subsection{Result Comparison}


The second experiment explained in Section \ref{pm} uses the same dataset as in the SOA \cite{chowdhury2019explaining}. Because of the unavailability of the feature extraction tool used in \cite{chowdhury2019explaining}, we could not reproduce the work. However, the datasets used in the SOA and our work are the same. Hence, we compare the results with those of the SOA paper. The results published in SOA paper \cite{chowdhury2019explaining} are based on 10\% of their dataset. On the contrary, we use 1,000 packets from each CoS traffic for training and the rest for testing. The exact number of test segments in our experiment is shown in Table \ref{table:Accuracy and FNR for dataset I}. Except for P2P and Email, our method achieves lower FNR for all other CoS types, and the overall accuracy is 99.13\%, whereas the SOA achieves only 92.5\% as shown in Table \ref{table:FNR comparision}. In our method, we observe that the CoS type, which uses asynchronous communication \cite{vonderwell2003examination} with more extensive delays, is not appropriately modelled. In our dataset, P2P and Email are asynchronous types of communication with massive delays. Chat is another asynchronous type, but there are no considerable delays in the dataset. Hence, in Table \ref{table:FNR comparision}, we can observe that chat is the third-worst performing class.

\section{Conclusion and Future Work}



We develop a novel segmented learning that includes the EVR and S2MC algorithms. We represent the CoS traffic in segments to model a traffic classifier. Then, S2MC determines the number of segments required for modelling to achieve superior classification performance. The proposed method is validated using multiple test scenarios that contain a number of real-world class-of-service network traffic flows. We employ multiple performance metrics to study the performance of the solution and show that it outperforms the state-of-the-art solution \cite{chowdhury2019explaining}. To the best of our knowledge, ours is the first work to classify more than 6 CoS traffic, including the latest hot trend VR traffic. Furthermore, our method exhibits exceptional results compared to the current solutions and stands as the state-of-the-art method for quick convergence, generalized, lightweight and robust CoS NTC.


Moving forward in the research, we need to investigate the modelling of asynchronous CoS types that introduce irregularities and abnormalities. A statistical variation imposes complexity in capturing the holistic characteristics of such asynchronous traffic. We need to test the solution in multiple networks to find the fruition of segmented learning in different conditions. We wish to explore incremental learning to improve the scalability of model and avoiding training from scratch to save time during deployment.


\bibliographystyle{IEEEtran}
\balance
\bibliography{IEEEabrv,refs}


\end{document}